\documentclass[conference]{IEEEtran}
\IEEEoverridecommandlockouts
\usepackage{cite}
\usepackage{amsmath,amssymb,amsfonts}
\usepackage{algorithmic}
\usepackage{graphicx}
\usepackage{textcomp}
\usepackage{xcolor}
\usepackage{pgfplots}
\usepackage{tabularx}
\usepackage{booktabs} 
\usepackage{array} 
\pgfplotsset{compat=1.17}
\usepackage{tikz}
\usetikzlibrary{arrows}
\def\BibTeX{{\rm B\kern-.05em{\sc i\kern-.025em b}\kern-.08em
    T\kern-.1667em\lower.7ex\hbox{E}\kern-.125emX}}
\begin{document}

\title{Entanglement Witness Derived By Using Kolmogorov-Arnold Networks }

\author{
    1\textsuperscript{st} Fatemeh Lajevardi\textsuperscript{1}, 
    2\textsuperscript{nd} Azam Mani\textsuperscript{1}, 
    3\textsuperscript{rd} Ali Fahim\textsuperscript{1} \\
    \textsuperscript{1}\textit{Department of Engineering Sciences, College of Engineering, University of Tehran, Tehran, Iran} \\
    Emails: fatemeh.lajevardi@ut.ac.ir, mani.azam@ut.ac.ir, a.fahim@ut.ac.ir
}

\maketitle

\begin{abstract}
We utilize Kolmogorov-Arnold Networks to design an interpretable model capable of detecting quantum entanglement within a set of nine-parameter two-qubit states. This network serves as an entanglement witness, achieving an accuracy of $94\%$ in distinguishing entangled states. Additionally, by analyzing the output functions of the KAN models, we explore the significance of each parameter (feature) in identifying the presence of entanglement. This analysis enables us to rank the features and eliminate the less significant ones, leading to the development of new entanglement witness functions that rely on fewer number of features, and hence do not require complete state tomography for their evaluation.
\end{abstract}


\section{Introduction}
Quantum entanglement is one of the most important features of quantum mechanics, underpinning many phenomena and technologies that distinguish quantum systems from their classical counterparts. It serves as a critical resource for quantum computation, quantum communication, and quantum cryptography, enabling applications such as quantum teleportation and super-dense coding \cite{nielsen2000quantum}. 
Entanglement is in fact a type of quantum correlation that can exist in a composite system of at least two parts. The Quantum state of a composite system is entangled if it can not be prepared solely by local operations and classical communications (LOCC) of the parties involved.
However, detecting and quantifying entanglement, especially in systems of increasing complexity, remain significant challenges in quantum information science \cite{vedral1997nphard, gharibian2010nphard}. Entanglement witnesses, a class of Hermitian operators, provide a practical solution for identifying entangled states from the on-entangled ones (separable stats) \cite{guhne2009entanglement}. By measuring the expectation value of a witness operator, one can determine whether a quantum state is entangled, i.e. a negative expectation value of an entanglement witness indicates the presence of entanglement. 
It is also challenging to define an appropriate witness operator to ensure positive results for separable states and negative results for most entangled states, that is why many researches have been done to design practically measurable entanglement witnesses for different system sizes \cite{horodecki1996separability, mohamad2004witnesses}. For example, for two-qubit systems, the positivity of the partial transpose (PPT) of the density matrix serves as a witness function \cite{asher1996ppt}, although it still requires complete tomography of the state.\\

Recently, the widespread adoption of machine learning and deep learning, along with their effective solutions, has led many researchers to embrace these methods for addressing the entanglement witness issue. For example, some entanglement witnesses are offered based on the SVM techniques, for two- and three-qubit systems \cite{sanavio2023svm, alexander2023svm, mahdian2024wernner}.\\

Kolmogorov-Arnold Networks (KANs) are a class of recently introduced neural networks which offer innovative alternatives to traditional multi-layer perceptrons (MLPs) in the field of neural networks \cite{Ziming2024kan}. The backbone of this network is based on the Kolmogorov-Arnold theorem, which states that for a multivariate continuous function on a bounded domain, it is possible to construct the function, by using a summation of continuous functions of single variables. This result can be explained by the following equation \cite{kolmogorov}.

\begin{equation}\label{eq:kolmogorov_equ}
f(x) = f(x_1, \ldots, x_n) = \sum_{q=1}^{2n+1} \Phi_q \left( \sum_{p=1}^n \phi_{q,p}(x_p) \right)
\end{equation}
where:
\begin{itemize}
	\item \( f : [0, 1]^n \to \mathbb{R} \)
    \item \( \phi_{q,p} : [0, 1] \to \mathbb{R} \) are continuous functions of single a variable,
    \item \( \Phi_q : \mathbb{R} \to \mathbb{R} \) are continuous functions.
\end{itemize}

This network is designed to learn the univariate functions \( \phi_{q,p} \) and \( \Phi_q \) instead of directly learning the multivariate function \( f \). This decomposition reduces the complexity of the problem, as learning univariate functions is often easier than learning multivariate ones. Unlike MLPs, which use fixed activation functions assigned to individual neurons, KANs employ learnable activation functions applied to the edges or weights of the network, replacing conventional linear weight matrices with spline functions \cite{Ziming2024kan}.{Figure \ref{fig:kan_structure} illustrates the schematic structure of such networks.
The unique architecture of KAN facilitates the effective summation of incoming signals while avoiding typical nonlinearities, potentially resulting in more compact computation graphs compared to their MLP counterparts. KANs harness the complementary strengths of both MLPs and splines: MLPs excel at learning compositional structures, while splines are adept at accurately approximating low-dimensional functions. Consequently, KANs effectively mitigate the curse of dimensionality associated with spline methods, enhancing both accuracy and interpretability across various applications, particularly in scientific domains. Empirical evaluations demonstrate that KANs consistently outperform MLPs, especially in small-scale artificial intelligence  and scientific tasks \cite{farhad2024kavvsmlp}. \\
\begin{figure}[t]
  \centering
  \includegraphics[width=\columnwidth]{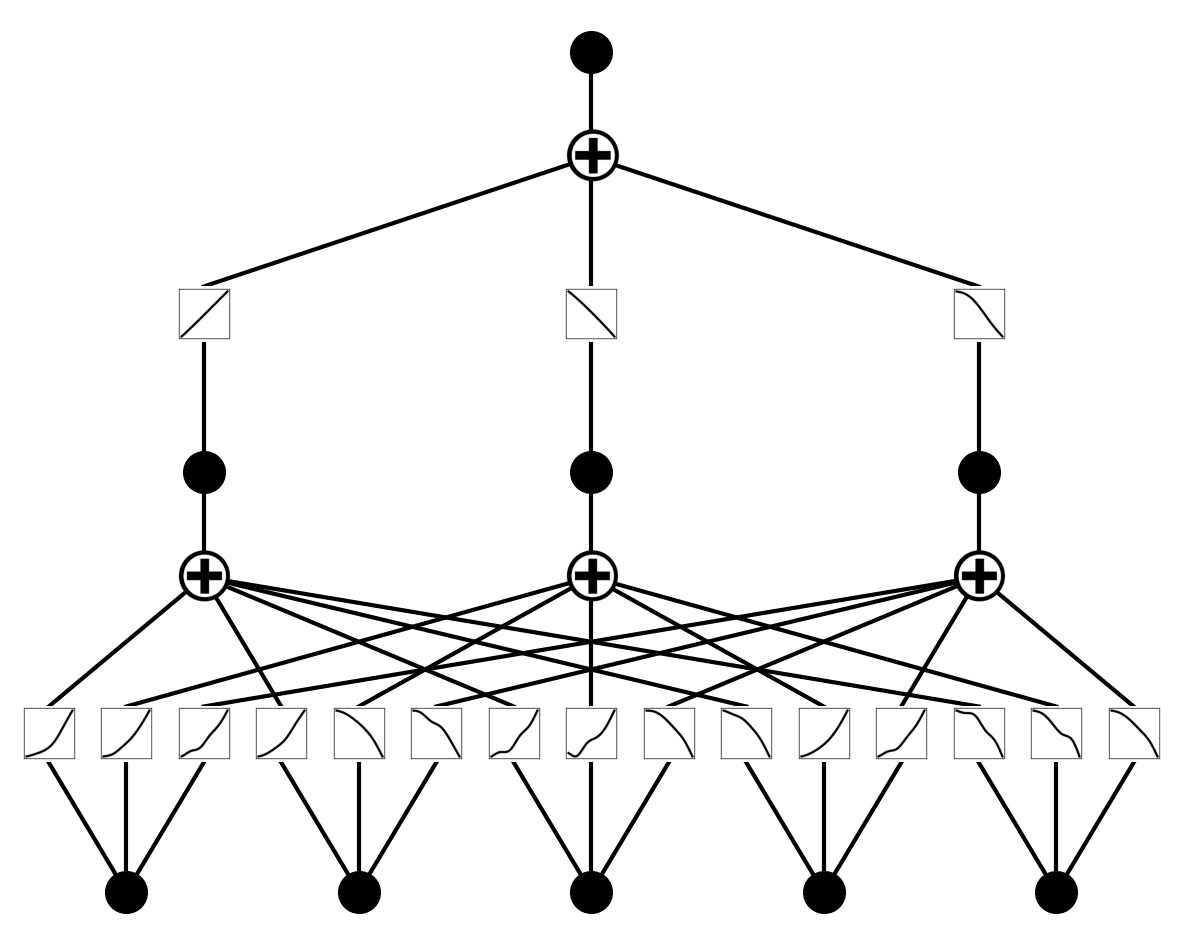} 
  \caption{
  The schematic structure of a Kolmogorov-Arnold Neural Network (KAN): This sample model consists of an input layer with 5 nodes, a single hidden layer containing 3 nodes, and an output layer with 1 node. As illustrated, each edge in the network is associated with an activation function that must be optimized during the learning process. The outputs of these activation functions are summed before being passed to the nodes of the subsequent layer.}
  \label{fig:kan_structure}
\end{figure}

Interpretability of the results provided by Kolmogorov-Arnold Networks, motivated us to introduce a novel method for detecting entanglement of two-qubit states by utilizing these Networks. Rooted in the Kolmogorov-Arnold representation theorem, KANs offer a robust framework for approximating functions while maintaining interpretability and accuracy. Our research focuses on two-qubit systems, and we construct a dataset of quantum states that spans both entangled and separable regions, ensuring uniform distribution across the data space. By training KAN models on this dataset, we demonstrate their capability to classify quantum states with high accuracy. Additionally, by utilizing the output functions of the KAN network, we derive some entanglement witnesses that do not necessitate complete state tomography for the evaluation of their expectation values, in other words, we reduce the required features.
There are also other studies that have developed simple networks to reduce the number of measurements needed for entanglement detection \cite{raeisi2024wernner}. However, it is important to note that the derived measurements from these methods are not locally applicable. While we decrease the number of required features, which directly leads to a reduction in the number of necessary local measurements.
On the other hand, we highlight that many studies examining the entanglement properties of two-qubit systems within the framework of machine learning typically focus on the one-parameter family of Werner states to construct their entangled datasets. In contrast, our research expands this scope by considering a nine-parameter family of both entangled and separable two-qubit states for our dataset, significantly broadening the range compared to previous investigations \cite{alexander2023svm, mahdian2024wernner}. \\

The paper is structured as follows: In section \ref{States} we briefly review the feature space of two qubit states and describe the structure of our data set. Section \ref{structure} details the architecture and training process of the KAN models, and section \ref{witness-extraction} discusses the extraction of entanglement witnesses and their validation. Finally, Section \ref{conclusion} presents the conclusions and potential avenues for future research.





\section{Two-Qubit Systems}\label{States}

Any two-qubit state, represented by a $4\times4$ density matrix, can be expanded in terms of Pauli matrices
\begin{equation}\label{general-rho}
	\rho=\frac{1}{4}\left[I\otimes I + \sum_{i=1}^{3} a_i \sigma_i \otimes I +  b_i I \otimes \sigma_i + \sum_{i,j=1}^{3} t_{ij} \sigma_i \otimes \sigma_j \right],
\end{equation} 
where $\otimes$ represents the tensor product, $\sigma_i,\ i=\{1,2,3\}$ stands for the Pauli matrices, $a_i$, $b_i$ and $t_{ij}$ are real coefficients. Hence, any arbitrary two qubit state has $15$ independent  features, that physically are the expectation values of the Pauli matrices and can be extracted by Pauli matrix measurements. For example, $t_{ij}$ is the expectation value of $\sigma_i \otimes \sigma_j$, i.e. $t_{ij}=Tr(\rho \ \sigma_i \otimes \sigma_j)$, and it can be obtained by measuring the hermitian observable $\sigma_i \otimes \sigma_j$.
Using representation (\ref{general-rho}) facilitates the future definition of entanglement witnesses that should be hermitian observables. We note that, while the Pauli matrices are traditionally denoted  as $\sigma_1$, $ \sigma_2 $, and $ \sigma_3 $, it is sometimes more convenient and clearer to adopt the notation $ X $, $ Y $, and $Z $ for these matrices, respectively. Accordingly, we will use the symbols $ XX $ and $XY $ in place of $ \sigma_1 \otimes \sigma_1 $ and $ \sigma_1 \otimes \sigma_2 $, respectively, applying similar compact notation for the other combinations. This notation will be used more throughout the section \ref{witness-extraction}.\\

Here, we only consider two-qubit states with maximally mixed marginals, i.e. $a_i=b_i=0$, for $i=\{1,2,3\}$. To generate such states, one can randomly generate sets of the coefficients $t_{ij}$ and only keeps the sets that ensure the positivity of density matrix (\ref{general-rho}). This approach is not only inefficient, but it also fails to yield a uniform distribution of density matrices as obtained through Haar measure.
\\

An alternative approach involves beginning with locally rotated versions of our target states, specifically the states referred to as \textit{X-states}. Once we generate random X-states, we can obtain arbitrary states with maximally mixed marginals by applying random local unitaries to these X-states. A general X-state can be written as 
\begin{equation}\label{X-state}
	\rho_{\text{x}} = \frac{1}{4} \left( I \otimes I + \sum_{k=1}^3 t_k \, \sigma_k \otimes \sigma_k \right),
\end{equation}
which has non-zero values only along the diagonal and off-diagonal elements, while representing in the computational basis. The positivity of the density matrix (\ref{X-state}), restricts the coefficients $t_1$, $t_2$, and $t_3$ to lie in a tetrahedron defined by the following equations
\begin{eqnarray}\label{tetra}
	 1 - t_1 - t_2 - t_3  & \geq 0, \cr
	 1 - t_1 + t_2 + t_3  &\geq 0, \cr
	 1 + t_1 - t_2 + t_3  &\geq 0, \cr
	 1 + t_1 + t_2 - t_3  &\geq 0.
\end{eqnarray} 

As mentioned above, to ensure the creation of randomly distributed two-qubit states with maximally mixed marginals, we must undertake two key actions. First, we need to generate uniform samples of X-states. Second, we should apply a collection of appropriately uniformly distributed unitary matrices, to locally rotate these X-states. It is important to note that since local unitary operators do not alter the physical properties of the systems, the source and target classes of states (Entangled or Separable) remain invariant, while their domain expands across the two-qubit state space. For the first step, we randomly generate the point $(t_1, t_2, t_3)$ within the constraints $ -1 \leq t_1, t_2, t_3 \leq 1$, and we include it in the X-state dataset if it satisfies equation (\ref{tetra}). For the second step, we generate random unitary operators $U_1$ and $U_2$ with accord to the Harr measure, and we add the following states to the data set
\begin{equation}\label{rho-9}
	\rho=(U_1 \otimes U_2) \rho_\text{x} (U_1^\dagger \otimes U_2^\dagger ).
\end{equation}
Uniformity of the states $\rho_\text{x}$, and the operators $U_1$ and $U_2$ warrants the uniformity of our data set. The states (\ref{rho-9}) were then rewritten in the form (\ref{general-rho}), to obtain nine independent features $t_{ij}$. \\

We finally used the PPT criterion to determine the class of each X-state (the label associated to each X-state). Once the labels of X-states are determined, the labels of other states will automatically be obtained, since the local unitary operators do not change the label of states.\\

In addition to the above dataset, we have also considered a family of  symmetric states as an extra dataset. These states are invariant under local unitary rotations around the z-axis, i.e. 
\begin{equation}
\left( U \otimes U \right) \rho \left( U^\dagger \otimes U^\dagger \right) = \rho, \hspace{0.5 cm} U=e^{i\theta \sigma_z},
\label{eq:symmetricState}
\end{equation}
where $\theta$ is an arbitrary real parameter. The above condition reduces the number of no-zero features from nine to five. The Positive Partial Transpose (PPT) criterion has also been utilized for labeling this set.

\section{Model Structure} \label{structure}
The desired network for processing the generated dataset is KAN. To design a model using KAN, the PyKan library, which is available on GitHub, was employed \cite{gitkan}. In this research, two models were generated, based on the type of our dataset. Each model utilized a dataset consisting of $100,000$ rows, $70\%$ of which was used for training, $20\%$ for validation, and $10\%$ for testing to assess the model’s accuracy. Half of the dataset consists of separable states, while the other half comprises entangled states.\\
The initial models are as follows: For the first dataset, we used a model with an architecture comprising two hidden layers with sizes $6$ and $3$, respectively, an input layer of size $9$, and an output layer of size $1$. For the second dataset, we developed a model with only one hidden layer of size $3$, an input layer of size $5$ and an output layer of size $1$. The number of the hidden layers and the size of each layer were considered as hyperparameters of the model and determined in the validation stage beside the other model's parameters.\\

To investigate the robustness of our models against the noise, we add some noise to the datasets in order to help the models adapt to noisy data (noisy measurement results). This noise was generated according to a Gaussian distribution with $0$ mean and a standard deviation of $0.1$. Considering this noise, the accuracy of both models were reduced to the reliable values $90\%$, from $94\%$ and $0.98\%$.

\begin{table}[ht]
\caption{Classification Report Model 9-6-3-1}
\label{tab:classification_report_9631}

	\centering
    \begin{tabular}{lcccc}
        \hline
        \textbf{} & \textbf{precision} & \textbf{recall} & \textbf{f1-score} \\
        \hline
        \textbf{Separable} & 0.95 & 0.94 & 0.94 \\
        \textbf{Entanglement} & 0.94 & 0.95 & 0.94 \\
        \hline
        \textbf{accuracy} &  &  & \textbf{0.94}  \\
        \hline
    \end{tabular}
\end{table}
\begin{table}[ht]
\caption{Classification Report Model 5-3-1}
\label{tab:classification_report_531}
	\centering
    \begin{tabular}{lcccc}
        \hline
        \textbf{} & \textbf{precision} & \textbf{recall} & \textbf{f1-score} \\
        \hline
        \textbf{Separable} & 0.96 & 1.00 & 0.98 \\
        \textbf{Entanglement} & 1.00 & 0.96 & 0.98 \\
        \hline
        \textbf{accuracy} &  &  & \textbf{0.98}  \\
        \hline
    \end{tabular}
\end{table}

\section{Extracting Witness From Model} \label{witness-extraction}

The defining characteristic of the Kolmogorov-Arnold Network that captures our attention is its capability to elucidate the model's final output as a function of the input features. In the context of the entanglement decision problem, a KAN generates a function that takes the initial features as inputs and produces an output value ranging from $0$ to $1$. If this output value is less than 0.5, the input state is classified as separable; otherwise, the state is categorized as entangled. Notably, this output function serves as an entanglement witness. For the models discussed in the previous section, the input features encompass all parameters of the density matrix, necessitating complete state tomography to determine the function's output. In this section, we only consider our general data set with nine features, and we analyze the output function to identify the most significant features relevant to the entanglement decision task. Our goal is to construct entanglement witness functions that utilize fewer features, thereby eliminating the need for complete state tomography. It's worth noting that our approach can also be utilized in the symmetric dataset, but we've left out that analysis for the sake of brevity.\\

Before proceeding, we need to clarify the notation that will be used for the remainder of the paper. As previously mentioned, our dataset consists of a family of quantum states characterized by nine features $t_{ij}$, where $i,j \in \{1,2,3\} $. Considering the expansion 
\begin{equation}\label{rho-9-2}
	\rho = \frac{1}{4}\left[I \otimes I + \sum_{i,j=1}^{3} t_{ij} \sigma_i \otimes \sigma_j \right],
\end{equation}
the $t_{ij}$ coefficients represent the expectation values of the Pauli observables $\sigma_i \otimes \sigma_j$. For example $t_{11}$ ($t_{23}$) is the expectation value of the operator $\sigma_1 \otimes \sigma_1$  ($\sigma_2 \otimes \sigma_3$), or by using the more convenient notation explained after equation (\ref{general-rho}), $t_{11}$ ($t_{23}$) is the expectation value of the operator $XX$  ($YZ$). Accordingly, when we talk about the importance of $XX$ ($YZ$) in the upcoming figures and tables, we are referring to the importance of the feature $t_{11}$ ($t_{23}$) that can be obtained from the $XX$ ($YZ$) measurement results. Similar conventions are also used for other features. Fixing this notation, we are now ready to present our method for constructing entanglement witness functions by using fewer features.  \\

After running our KAN model for a dataset of $100,000$ density matrices, we analyzed the importance of each feature in the value of the output function. The output function is composed of a combination of sinusoidal and linear functions of features equation (\ref{eq:witness_function_1}) shows a  sample of such function. Although the PyKan library provides options to select various functions for model explanations, we opted for these two in order to more effectively assess feature importance. Figure \ref{fig:coeff_observables} represents this analysis for the exported function.  It is evident from the figure that the feature $ZX$ has the minimal impact on the classification of the states, while the features $XZ$, $YZ$, and $ZZ$ are identified as the most influential features in this specific model.

\begin{figure}[ht]
\centering
\begin{tikzpicture}
    \begin{axis}[
        width=0.9\columnwidth, 
        height=5cm,
        ybar,
        ymin=0,
        ymax=0.35,
        bar width=7pt, 
        symbolic x coords={XX, XY, XZ, YX, YY, YZ, ZX, ZY, ZZ},
        xtick=data,
        xlabel={Observables},
        ylabel={Values},
        title={Coefficient of Observables},
        xticklabel style={font=\small, rotate=45, anchor=east}, 
        yticklabel style={font=\small},
        every axis title/.append style={font=\small},
        grid=major,
        nodes near coords,
        every node near coord/.append style={font=\scriptsize}, 
    ]
    \addplot+[ybar,fill=blue,opacity=0.7] coordinates {
        (XX, 0.05)
        (XY, 0.16)
        (XZ, 0.28)
        (YX, 0.13)
        (YY, 0.19)
        (YZ, 0.23)
        (ZX, 0.00)
        (ZY, 0.16)
        (ZZ, 0.20)
    };
    \end{axis}
\end{tikzpicture}
\caption{The coefficients of observables for a sample output function from the model 9--6--3--1 are displayed, with the y-axis representing the coefficient values of the corresponding features on the x-axis.}
\label{fig:coeff_observables}
\end{figure}
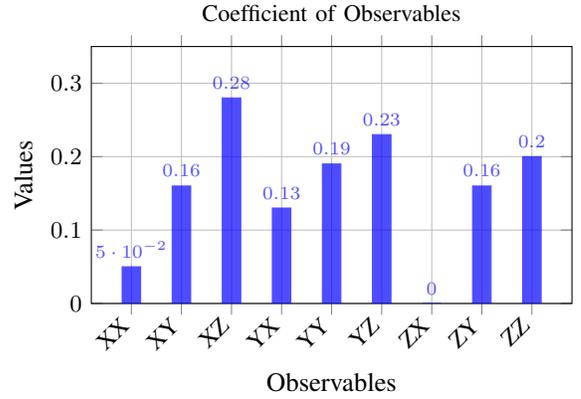

However, this analysis pertains to a single model, when multiple models are trained and their respective functions are analyzed, the conclusions may differ. To obtain more robust and reliable results, it is necessary to aggregate analyses across many models. To achieve this, we use a bootstrap method: we generate multiple independent datasets, apply a KAN model to each dataset, perform an analysis similar to that shown in Figure (\ref{fig:coeff_observables}) for each model, and then combine the results from all these models. Finally, we use a specific procedure (explained below) to assess the importance of each feature. Our analysis shows that considering $20$ independent datasets of $100,000$ density matrices and evaluating the $20$ output functions from these models can yield robust and reliable results.\\

After training $20$ models, in order to sort the features according to their importance in the output functions, for each model, we define eight groups for the features; namely Top-1, Top-2, ..., and Top-8. Top-1 contains the most important feature of the model, Top-2 contains the two most important features, and finally Top-8 contains all features except the less important one. 
We indicated the members of each group, for each model, and we then combined these individual results in Figure \ref{fig:top_ranking_chart}. The Figure shows each group by a line, and also illustrates the number of repetition of any feature as a members of each group. As a result, if we want to construct an entanglement witness, which depends only on four features (for example), then we should go to the Top-4 group, and select the four observables with the highest frequency of occurrence. We then construct a KAN model that works only with these four inputs. The output function of this KAN model will serve as an entanglement witness function. 
Table  \ref{tab:measurement_details} summarizes the models that can be constructed by the results obtained from Figure \ref{fig:top_ranking_chart}. \\

\begin{table}[ht]
    \centering
    \caption{Various degrees of feature reduction, along with their associated observables and model structure. Each observable was selected based on its highest frequency of appearance in Figure\ref{fig:top_ranking_chart}.}
    \begin{tabular}{c p{3cm} c} 
        \toprule
        \textbf{\#Measurements} & \textbf{Observables} & \textbf{Model} \\
        \midrule
        1 & YY & 1-3-1 \\
        2 & XZ, YY & 2-3-1 \\
        3 & XZ, YX, YY & 3-2-1 \\
        4 & XY, XZ, YX, YY & 4-2-1 \\
        5 & XX, XY, XZ, YX, YY & 5-3-1 \\
        6 & XX, XY, XZ, YX, YY, YZ & 6-3-1 \\
        7 & XX, XY, XZ, YX, YY, YZ, ZZ & 7-5-3-1 \\
        8 & XX, XY, XZ, YX, YY, YZ, ZX, ZY & 8-5-3-1 \\
        \bottomrule
    \end{tabular}
    \label{tab:measurement_details}
\end{table}

This approach helps us to construct entanglement witness functions that depend on arbitrary number of observables. 
However, it should be noted that reducing the number of observables also decreases the accuracy of the witnesses. Figure  \ref{fig:horizontal_bar_chart} illustrates the model's accuracy for different levels of feature reduction. This chart illustrates that using only four features (four observable measurements) enables us to classify states with over $80\%$ of accuracy. Since four measurements provide us with a satisfactory accuracy, bellow we will present the witness function that we have obtained from our $4-2-1$ model. Needless to say, this method is applicable to all proposed groups of observables presented in Figure \ref{fig:top_ranking_chart}.

\begin{figure}[th]
    \centering
    \begin{tikzpicture}
        \begin{axis}[
            width=\columnwidth, 
            height=8cm,         
            xlabel={\textbf{Features}},
            ylabel={\textbf{Frequency of Occurrence}},
            grid=major,
            ymin=0,
            ymax=22,
            xtick=data,
            symbolic x coords={XX, XY, XZ, YX, YY, YZ, ZX, ZY, ZZ},
            legend style={font=\footnotesize, at={(0.5,-0.2)}, anchor=north, legend columns=3},
            xticklabel style={rotate=45, anchor=east, font=\scriptsize},
            yticklabel style={font=\scriptsize},
            cycle list name=color list
        ]
        
        \addplot+[thick] coordinates {(XX, 3) (XY, 2) (XZ, 3) (YX, 2) (YY, 5) (YZ, 0) (ZX, 0) (ZY, 2) (ZZ, 3)};
        \addlegendentry{\scriptsize Top-1 \ }

        \addplot+[thick] coordinates {(XX, 6) (XY, 6) (XZ, 7) (YX, 6) (YY, 7) (YZ, 2) (ZX, 0) (ZY, 2) (ZZ, 4)};
        \addlegendentry{\scriptsize Top-2 \ }

        \addplot+[thick] coordinates {(XX, 7) (XY, 8) (XZ, 11) (YX, 8) (YY, 10) (YZ, 5) (ZX, 2) (ZY, 3) (ZZ, 6)};
        \addlegendentry{\scriptsize Top-3 \ }

        \addplot+[thick] coordinates {(XX, 10) (XY, 11) (XZ, 12) (YX, 13) (YY, 13) (YZ, 7) (ZX, 2) (ZY, 5) (ZZ, 7)};
        \addlegendentry{\scriptsize Top-4 \ }
        
        \addplot+[thick] coordinates {(XX, 14) (XY, 14) (XZ, 13) (YX, 13) (YY, 16) (YZ, 13) (ZX, 5) (ZY, 5) (ZZ, 7)};
        \addlegendentry{\scriptsize Top-5 \ }

        \addplot+[thick] coordinates {(XX, 16) (XY, 15) (XZ, 16) (YX, 14) (YY, 17) (YZ, 18) (ZX, 7) (ZY, 7) (ZZ, 10)};
        \addlegendentry{\scriptsize Top-6 \ }

        \addplot+[thick] coordinates {(XX, 18) (XY, 16) (XZ, 18) (YX, 15) (YY, 20) (YZ, 18) (ZX, 11) (ZY, 12) (ZZ, 12)};
        \addlegendentry{\scriptsize Top-7 \ }
        
        \addplot+[thick] coordinates {(XX, 19) (XY, 17) (XZ, 19) (YX, 16) (YY, 20) (YZ, 20) (ZX, 17) (ZY, 17) (ZZ, 15)};
        \addlegendentry{\scriptsize Top-8 \ }
        
        \end{axis}
    \end{tikzpicture}
    \caption{Display the importance of observables in each group, where each line represents a group. For observables on the X-axis, the Y-axis indicates their frequency of occurrence within that group. For example, if a witness requires three observables, refer to the 'Top 3' line and select the three observables with the highest frequencies indicated on the Y-axis. }
    \label{fig:top_ranking_chart}
\end{figure}
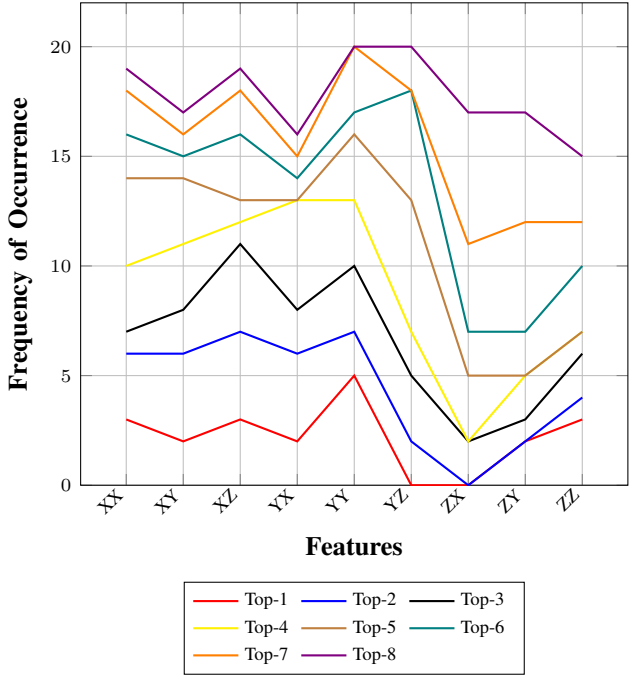

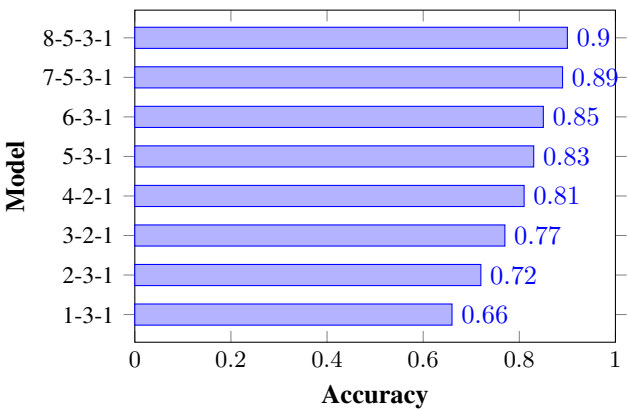
\begin{figure}[th]
    \centering
    \begin{tikzpicture}
        \begin{axis}[
            xbar,                                  
            width=0.9\columnwidth,                   
            height=6cm,                            
            xlabel={\textbf{Accuracy}},              
            ylabel={\textbf{Model}},            
            xmin=0, xmax=1,                        
            symbolic y coords={1-3-1, 2-3-1, 3-2-1, 4-2-1, 5-3-1, 6-3-1, 7-5-3-1, 8-5-3-1},
            ytick=data,                            
            nodes near coords,                     
            nodes near coords align={horizontal},  
            bar width=8pt,                         
            yticklabel style={align=right, font=\small},  
            xticklabel style={font=\small}         
        ]
        
        \addplot coordinates {
            (0.66,1-3-1) 
            (0.72,2-3-1)
            (0.77,3-2-1) 
            (0.81,4-2-1) 
            (0.83,5-3-1)
            (0.85,6-3-1)
            (0.89,7-5-3-1)
            (0.90,8-5-3-1)};
        \end{axis}
    \end{tikzpicture}
    \caption{Accuracy of each models with different number of reduced feature. All potential size of features, ranging from 1 to 8, are inserted.}
    \label{fig:horizontal_bar_chart}
\end{figure}

For the four observables \( YX \), \( YY \), \( XZ \), and \( XY \), the extraction function phase concluded after training the model with these features 
Subsequently, the \texttt{auto\_symbolic} function from the PyKan library was used to fit a curve to the output generated by KAN. This function proposed sinusoidal and linear combinations of observables as witness functions. Equation (\ref{eq:witness_function_1}) shows a  sample witness function for this model. To simplify the witness and improve the readability, coefficients are rounded to two decimal places. It is valuable to note that one can obtain other equivalently important groups of observables and hence entanglement witnesses by cyclic permutations of Pauli matrices $X-Y-Z$.  

\begin{flalign}
    \label{eq:witness_function_1}
    W &= -0.12 \sin(5.49YX + 1.61) - 0.08 \sin(4.27YX - 7.79) \notag \\
    &+ 0.17 \sin(3.15XY - 1.58) - 0.15 \sin(4.43XY - 1.6) \notag \\
    &+ 0.14 \sin(5.02XZ - 1.57) - 0.06 \sin(4.41XZ + 1.6) \notag \\
    &+ 0.18 \sin(4.24YY - 1.59) + 0.22 \sin(2.65YY - 1.57)\notag \\
    &+ 1.88
\end{flalign}



\section*{Discussion} \label{conclusion}

The results of this research demonstrate the potential of Kolmogorov-Arnold Networks (KANs) in the development of quantum entanglement witnesses for two-qubit systems. By leveraging the interpretability and flexibility of KANs, we successfully identified entanglement with a high degree of accuracy, using minimal measurement resources. 
Traditional methods for detecting entanglement often require complete state tomography or non-local measurements, which are resource-intensive. By contrast, the KAN-based approach presented in this study reduces the number of required measurements while maintaining high accuracy. This feature is particularly advantageous in experimental quantum setups, where measurement costs and noise can pose significant challenges.
When comparing our KAN model with deep learning models such as \cite{raeisi2024wernner, unsupervised2021Yiwei, DL2024Asif}, there may be no significant advantage in accuracy, as these models achieve near-perfect performance. However, in contrast to our model, their complex and non-descriptive nature makes it impossible to extract entanglement witnesses. On the other hand, SVM-based methods are more interpretable, allowing for witness extraction, but the key consideration of these models (compared to ours) is the restricted family of entangled states analyzed \cite{sanavio2023svm, alexander2023svm, mahdian2024wernner}. For example, \cite{mahdian2024wernner} focuses only on the Werner states, and \cite{alexander2023svm} examines just a subset of this family, achieving nearly 100\% accuracy for these specific families. However, \cite{sanavio2023svm}, which evaluates all entangled states, achieves 92\% accuracy, lower than the 94\% accuracy of our proposed method. In summary, the proposed method seeks to integrate the strengths of different approaches, but this comes at the cost of not delivering perfect accuracy across all cases.

While the proposed method shows significant promise, it is currently restricted to two-qubit systems. Extending the approach to multi-qubit and higher-dimensional systems, is the goal of our future research. As system size increases, the complexity of the entanglement structure grows exponentially, potentially requiring modifications to the KAN architecture or optimization techniques.

\section*{AI Usage Statement} \label{AIUsage}
The authors declare that they used artificial intelligence tools only to improve and edit the text of the article.

\end{document}